\documentclass[english]{article}
\usepackage[utf8]{inputenc}
\usepackage{amstext}
\usepackage{amssymb}
\usepackage{graphicx}
\usepackage{esint}

\newcommand{\el}{\ell}

\newcommand{\pa}{\partial}

\newcommand{\ug}{ \; = \; }

\newcommand{\bb}{\begin{equation}}
\newcommand{\ee}{\end{equation}}
\newcommand{\bega}{\begin{eqnarray}}
\newcommand{\ega}{\end{eqnarray}}
\newcommand{\begae}{\begin{eqnarray*}}
\newcommand{\egae}{\end{eqnarray*}}

\newcommand{\h}{\hspace*{4ex}}
\newcommand{\dis}{\displaystyle}

\newcommand{\om}{\omega}

\newcommand{\zetam}{\zeta_m}
\newcommand{\zetarm}{\zeta_{r\,m}}
\newcommand{\zetaim}{\zeta_{i\,m}}

\makeatletter

\newcommand{\address}[1]{
\par {\raggedright #1
\vspace{1.4em} \noindent\par} }

\usepackage{cite}

\makeatother

\usepackage{babel}

\begin{document}

\baselineskip 0.55cm

\title{Fusion Between Frozen-Wave-Type Beams and Airy-Type
Pulses: Diffraction-Dispersion-Attenuation Resistant Vortex Pulses in
Absorbing Media}


\author{Michel Zamboni-Rached$^{1}$\footnote{E-mail address for
contacts: mzamboni@decom.fee.unicamp.br} and Mo Mojahedi$^{2}$}

\maketitle

\address{$^{1}$University of Campinas, Campinas, SP, Brazil.

$^{2}$Department of Electrical and Computer Engineering at the
University of Toronto, Toronto, ON, Canada.}

\begin{abstract} In this paper we perform a fusion between
two important theoretical methodologies, one related to the Frozen Wave beams, which are non-diffracting beams whose longitudinal
intensity pattern can be chosen \emph{a priori} in
an medium (absorbing or not), and the other related to the Airy-Type pulses, which
are pulses resistant to dispersion effects in dispersive materials. As a result, a new method emerges, capable of
providing vortex pulses resistant to three concomitant effects, i.e.: diffraction, dispersion and attenuation; while concurrently the spatial variation of the wave intensity
along its axis of propagation can be engineered at will. The new approach
can be seen as a generalization of the Localized Waves theory in the paraxial regime and
the new pulses can have potential applications in
different fields such as optics communications, nonlinear optics, micromanipulation, and so on.

\end{abstract}

\section{Introduction}

\h Since the works of Sheppard and Wilson \cite{sheppard},
Brittingham \cite{brit} and Durnin \cite{durnin}, the subject of
non-diffracting beams and pulses, also known as Localized Waves,
has been of interest to many researches. It is now a well
established fact that non-diffracting waves \cite{livro1,livro2}
such as the Bessel beams can resist diffraction effects for long
distances, when compared to the ordinary waves. Moreover, in
Ref.\cite{fw1} it was shown that it is possible to model the
longitudinal intensity pattern of the non-diffracting beams at
will, in an approach dubbed \emph{Frozen Waves} (FWs).
Such intensity pattern, chosen a priori, can be constructed on
axis ($\rho = 0$) or on the surface of a cylinder \cite{fw3} of radius
$\rho = \rho_{\nu}$, where the resulting beams -- called Frozen Waves
(FWs) -- were experimentally demonstrated in Refs. \cite{tarcio1,tarcio2}. Later, the FW
method was extended in \cite{fw2,ahmed2}, allowing the spatial shaping of
non-diffracting beams to take place in absorbing materials. That was an
important step forward in the Localized Wave theory, as it
provided beams not only resistant to the diffraction but
also to the attenuation. The theoretical finite energy version of
diffraction-attenuation resistant beams was developed in \cite{fwbg,fwmicro}, and later on it was shown (theoretically and experimentally) that FW beams can be used
to control the orbital angular momentum (OAM) \cite{ahmed1} and the polarization \cite{mateus1,mateus2} of a beam
along its axis of propagation. In short, today, FWs can be understood as a class of structured light in which the beam's intensity pattern, polarization, OAM (magnitude and sign), and wavelength can be engineered, almost at will, along its axis of propagation.

\h As related to pulses and their resistance to diffraction, they have been discussed in Refs. [2,17-30], where the theory was later extended to include pulses resistant to both diffraction and dispersion [31-38]. However, in order to adorn the above mentioned pulses with a degree of resistance toward dispersion, a complicated space-time
coupling in their spectra was needed. But, with the advent of Airy pulses the situation became relatively simpler since such pulses only required a suitable cubic phase in the frequency spectrum [39-44]

\h In summary, today, within the Localized Wave theory, we have: i)
beams that can resist diffraction and attenuation effects when
propagating in unguided absorbing media, and ii) pulses resistant
to the diffraction and dispersion effects in unguided dispersive
media. In light of our previous discussion regarding FWs and structured light,
naturally, a question can be asked: is it possible to engineer a vortex pulse
that is concurrently resistant to
the three concomitant effects of diffraction, dispersion and
attenuation in unguided dispersive absorbing media, while at same time --
similar to the case of FWs -- the pulse intensity can display priorly chosen
values at different locations (regions) along its direction of propagation?
Undoubtedly, demonstrating such capability can open new venues for other
possibilities such as engineering the pulse polarization, (local) OAM and central
wavelength along the propagation. However, these particular considerations will
be postponed to future communications.

\h In this paper, we show that it is possible to construct pulses that, over a
finite distance, are immune to diffraction, dispersion and attenuation while, at
same time, their intensities can assume arbitrary values and patterns along the
propagation. The paper is organized as the following. In section 2, we develop
the required theoretical frame work, whereas section 3 describes some of the
applicable examples. Section 4 contains our final remarks and conclusions.
We expect that the theoretical formulation presented here will have an important
impact in multiple areas of research in optical sciences such as optical
communications, non-linear optics and optical manipulations.

\section{The Method}

\h Let us first present the basic equations that describe the
evolution of a pulse in a linear media. A pulse, $\Psi(\mathbf{r},t)$,  with slowly varying envelope, propagating in a dispersive absorbing medium with complex index of refraction,
$n(\om) = n_R(\om) + i n_I(\om)$, can be
described as:

\bb \Psi(\mathbf{r},t) \ug e^{i k(\om_0) z}e^{-i \om_0 t}
A(\mathbf{r},t), \label{psi} \ee

with

\bb k(\om_0) \ug n_R(\om_0)\frac{\om_0}{c} + i
n_I(\om_0)\frac{\om_0}{c} = k_R(\om_0) + ik_I(\om_0) \equiv k_{R
0} + i k_{I 0}\ . \ee

The pulse envelope, $A(\mathbf{r},t)$, obeys

\begin{equation}
\frac{1}{2k(\om_0)}\nabla^2_{\perp} A + i\left(\frac{\partial}{\partial z} +
\beta_1\frac{\partial}{\partial t}\right)A -
\frac{\beta_{2}}{2}\frac{\partial^{2}A}{\partial t^{2}} \ug 0,
\label{eq3D}
\end{equation}

where

\bb \beta_1 \ug \frac{\pa k_R}{\pa \om}|_{\om_0} \,\,\,\,\, {\rm
and} \,\,\,\,\, \beta_2 \ug \frac{\pa^2 k_R}{\pa \om^2}|_{\om_0}
\,\,. \ee

\

\h By making the usual transformations, $z=z$ and $T = t-z/v_g$,
with $v_g = 1/\beta_1$, we can rewrite Eq.(\ref{eq3D}) as

\begin{equation}
\frac{1}{2k(\om_0)}\nabla^2_{\perp} A + i\frac{\partial}{\partial z}A -
\frac{\beta_{2}}{2}\frac{\partial^{2}A}{\partial T^{2}}\ug 0\ .
\label{eq3D2}
\end{equation}

\

\h Now, let us consider the pulse envelope, $A(\mathbf{r},t)$, having the
following form

\bb A(\mathbf{r},t) \ug  W(\rho,\phi,z)P(z,T) \,\, . \label{A}\ee

By substituting Eq.(\ref{A}) in (\ref{eq3D2}), we have

 \bb \frac{1}{2 k(\om_0)}\nabla^2_{\perp} W + i\, \frac{\pa W}{\pa z} \ug 0,  \label{W} \ee

and

 \bb i\,\frac{\pa P}{\pa z} - \frac{\beta_2}{2}\frac{\pa^2 P}{\pa T^2} \ug
 0\ . \label{P}  \ee

At this point it is important to note that $W(\rho,\phi,z)$,  governed by Eq.(\ref{W}),
 is the partial differential equation for the envelope of a
paraxial beam; while $P(z,T)$,  governed by Eq.(\ref{P}),  is
the partial differential equation for the envelope of a 1D pulse propagating in a
dispersive medium.

\h Having established the differential equations governing the behavior of  $W(\rho,\phi,z)$ and  $P(z,T)$, in the next subsection we will develop a space-time modeling of vortex pulses (i.e. pulses carrying OAM) propagating in  unguided, dispersive, and absorbing medium. Consequently, we shall see that it is possible to construct vortex pulses which resist the effects of dispersion, diffraction, and attenuation as they propagate in an unbounded medium. Such construction and formulation can be viewed as the generalization of the Localized Wave theory in the paraxial regime.

\subsection{Space-time Modeling of Diffraction, Dispersion, and Attenuation
Resistant Vortex Pulses}

\h The basic approach in developing vortex pulses, $A(\mathbf{r},t)$, that are immune to diffraction, attenuation, and dispersion is to enforce resistance to  diffraction and attenuation (along with control of the OAM) through the spatial function $W(\rho,\phi,z)$ and its corresponding differential equation, Eq.(\ref{W}); whereas to make the vortex pulses resistant to dispersion through the temporal function, $P(z,T)$, and its corresponding differential equation, Eq.(\ref{P}).

\h Similar to the approach in \cite{fwbg}, we choose  $W(\rho,\phi,z)$ to be a superposition of $2N+1$ copropagating $\nu$-order Bessel-Gauss beams given by

\begin{equation}
        W(\rho,\phi,z) = \frac{\exp\left(\dis{-q^2\frac{\rho^2}{\mu}}\right)}{\mu}
        \exp\left(\dis{- i k(\om_0) \frac{z}{\mu}}\right)\sum^{N}_{m=-N}A_{m}
        J_{\nu}\left(\eta_m\frac{\rho}{\mu}\right)e^{i\nu\phi}\,
        \exp\left(\dis{i \zeta_m \frac{z}{\mu}}\right), \label{FWbg}
        \end{equation}

where

 \bb \mu \ug 1 + i 2 \frac{q^2}{k(\om_0)} z \,\,. \label{mu} \ee

In Eq.(\ref{FWbg}), $q$ is a constant,  $A_m$ are coefficients of expansions to be determined, and $\eta_m$ and $\zeta_m$ are
the transverse and longitudinal wavenumbers (respectively) of the
mth Bessel-Gauss beam  in the superposition that must satisfy
$\zeta_m = k(\om_0) - \eta_m^2/2k(\om_0)$.

\h The solution (\ref{FWbg}) can be used
for obtaining a light beam, given by $ \exp(-k_{I 0}z)\exp(i
k_{R 0} z - i \om_0 t)W(\rho,\phi,z)$, resistant to the diffraction
and attenuation effects in an absorbing medium, with a longitudinal
intensity pattern that can be chosen \emph{a priori}. This intensity
pattern is given by a function, $|F(z)|^2$, of our choice,
and can be concentrated over the z axis (in the case $\nu=0$), with a
beam spot radius $r_0$, or over a cylindrical surface (in the case $|\nu| \geq 1$) of radius
$\rho_{\nu}$, both also of our choice, as we are going to explain soon. More specifically, one can have\footnote{Here, $\rho_0 \equiv 0$.} $\exp(-2k_{I
0}z)|W(\rho=\rho_{\nu},\phi,z)|^2 \approx |F(z)|^2$ within a
predefined longitudinal range $0 \leq z \leq L/2$, and
$\exp(-2k_{I 0}z)|W(\rho=\rho_{\nu},\phi,z)|^2 \approx 0$ for $0>z>L/2$.

\h It is important to say that while in the FW method $W$ shapes the beam envelope, here, due to Eqs.(\ref{psi},\ref{A}), such modeling takes place on the pulse, i.e., $|F(z)|^ 2$ becomes the pulse's peak intensity pattern along the propagation, the same occurring for the transverse features, i.e., $ r_0 $ becomes the pulse's spot radius (when $ \nu = 0 $) and $ \rho_ {\nu} $ becomes the radius of the vortex pulse, which has a donut-shaped profile.

For such modelling, the finite energy FW method \cite{fwbg} requires the following choices
for $\eta_m$ in the solution (\ref{FWbg}):

\bb \eta_m \ug \sqrt{2}\dis{\sqrt{1 - \frac{1}{k_{R
0}}\left(Q+\frac{2\pi m}{L} \right) }} \, |k(\om_0)| \,\, ,
\label{etan} \ee

which implies $\zetam = \zetarm + i\, \zetaim$, with

 \bb \left\{\begin{array}{l}
   \zetarm \ug Q + \dis{\frac{2 \pi m}{L}} \\
 \\
\zetaim \ug k_{I 0} \dis{\left(2 - \frac{\zeta_{r\,m}}{k_{R 0}}
\right)} \,\, ,
\end{array} \right. \label{zetam}
 \ee

where $Q$ is a positive constant, obeying $ 0 \leq Q + 2\pi N/L \leq k_{R 0} $
and being related
to the transverse dimensions of the beam. Actually, when $\nu=0$, the pulse spot
radius, $r_0$, will be approximately given by $r_0 \approx 2.4/\eta_0$ and,
when $|\nu| \geq 1$, the radius of the donut-shaped vortex pulse will be
approximately given by $\rho_{\nu}$, which corresponds to the first positive
root of $[(d/d\rho)J_{\nu}(\eta_0 \rho)]|_{\rho=\rho_{\nu}} = 0$.

Yet, according to \cite{fwbg}, the values of $q$ and of the coefficients $A_m$
have to be given by


\bb q \ug \frac{2 k_{R 0}}{L \, \eta_0} \label{q2}    \ee

and

\bb  A_{m} = \frac{1}{L} \int^{L}_{0}\frac{F(z)}{G(z)}\,
e^{-i\frac{2\pi}{L}m z} dz  \,\, , \label{An2} \ee

with

\bb G(z) \ug e^{- k_{I 0} z}\exp\left(\frac{-\eta_0^2(k_{I 0} + 2
q^2 z)z}{2 [k_{R 0}^2 + (k_{I 0} + 2 q^2 z)^2]}\right) I(z)
\label{G} \ee

where $I(z)$\footnote{The function $I(z)$ does not appear in the work of the
finite energy FWs \cite{fwbg} because there the method deals only with beams,
that is, it considers $P (z, T) = 1$.} is a function that is related, as we
are going to
see, with a possible intensity decay of $P(z,T)$.

\h So, by considering $W(\rho,\phi,z)$ given by Eq.(\ref{FWbg}), with
Eqs.(\ref{mu}-\ref{G}), we are ensuring that our 3D pulse solution
given by Eqs.(\ref{psi},\ref{A}): a) can be spatially modelled on demand,
presenting,
as a subproduct, resistance to the diffraction and attenuation
effects for long distances; b) it is endowed with OAM, being its topological
charge equal
to $\nu$. The distances of diffraction and attenuation resistance, that we
call $Z_{diff}$ and $Z_{att}$,
respectively, will be set by the morphological function $F(z)$ and so, as
that function is set to be null
for $z>L/2$, they can be considered as $Z_{diff} = Z_{att} = L/2$.

\h Here, we have to stress an important point. While the diffraction resistance
distance can be made arbitrarily large (even experimentally it can be made
hundred of times greater than the Rayleigh distance, $L_{diff}\approx \sqrt{3}\,k_{R0} r_0^2$),
the same does not occur with the attenuation resistance distance which, according
to the FW method \cite{fw2}, should not exceed about 10 times the penetration
depth $L_{att} = 1/\alpha$ (where $\alpha = 2 n_I \om /c $ is the absorption
coefficient), otherwise the lateral lobules can acquire very high intensity
levels. Due to this, the parameter $L$, used to set up the range $0 \leq z \leq L/2$
where the morphological function $F(z)$ is non-null, is limited to values about $10/\alpha$.

\h Now, let us move on and make a choice to $P(z,T)$, a solution of
(\ref{P}) which, we expect, shall provide resistance to the
dispersion effects to the resulting 3D pulse envelope.
For this purpose, a natural choice is a 1D Airy-type pulse,
as the finite energy Airy-exponential one given by


\begin{eqnarray}
P\left(z,T\right) & = &
\exp\left[\frac{6a\left(2\varepsilon\tau-Z^{2}\right)+
iZ\left(-6a^{2}-6\varepsilon\tau+Z^{2}\right)}{12}\right]\nonumber \\
 & \times & \mathrm{Ai}\left(\varepsilon\tau-\frac{Z^{2}}{4}-iaZ\right)\,\, , \label{airyfinito}
\end{eqnarray}

where $\mathrm{Ai}(.)$ is the Airy function, $a>0$ is a constant related
with the time exponential apodization and $\varepsilon = \pm 1$ determines the
direction of the Airy pulse envelope \cite{airycausal}, which can have the smaller peaks preceding
the main lobe ($\varepsilon = 1$) or the main lobe preceding the smaller
peaks ($\varepsilon=-1$). In Eq.(\ref{airyfinito}), we have used the normalized variables $\tau=T/T_{0}$ and
$Z=z\beta_{2}/T_{0}^{2}$, with $T_0$ a constant equal to the
initial time width of the main peak of the 1D pulse $P(z,T)$
which, depending on the value of the parameter $a$, can resist to
the dispersion effects for distances much longer than the usual
dispersion distance, $L_{disp} = T_0^2 / |\beta_2|$, of ordinary
gaussian pulses. A conservative estimation of the field
depth, $Z_{disp}$, of the Airy-Exponential pulse (\ref{airyfinito})
is given by

\bb Z_{disp} \ug \sqrt{\frac{2}{a}}\,\frac{T_0^2}{|\beta_2|} \ug
\sqrt{\frac{2}{a}} L_{disp} \,\,\, . \label{a}\ee

\h Although the finite energy 1D pulse solution exhibits resistance to the
dispersion effects, it also exhibits a continuous intensity decay from $z=0$.
For distances smaller than $Z_{disp}$, it is not difficult to show that the
pulse's peak intensity decays approximately according to the function

\bb I(z) \ug \exp\left(-\frac{a\beta_2^2}{4T_0^4}z^2 \right) \,\,. \label{I} \ee

The intensity decay given by (\ref{I}) is taken into account (i.e., it is
compensated) by $W(\rho,\phi,z)$, Eq.(\ref{FWbg}), through the coefficients
$A_m$ given by Eqs.(\ref{An2},\ref{G}).


\h In this way, with $P(z,T)$ given by Eq.(\ref{airyfinito}), we are ensuring
that our 3D pulse, Eqs.(\ref{psi},\ref{A}), is also resistant to the dispersion
effects for long distances.

\h At this point, an important observation has to be made about the 1D pulse
solution given by Eq.(\ref{airyfinito}). For causality reasons, which are very
well explained in \cite{airycausal}, in the case $\varepsilon = -1$, expression
(\ref{airyfinito}) should not be considered for $z \geq 2T_0^3/\beta_2^2 v_g$.
In our case, however, such limitation is not alarming at all because, according
to our method, the morphological function $F(z)$ ensures that the resulting pulse
will possess negligible intensities for $z \geq L/2$ which, in general, will be
set to values much smaller than $2T_0^3/\beta_2^2 v_g$.

\h The final result is that the resulting 3D pulse,
$\Psi(\rho,\phi,z,t)$, is given by

\begin{eqnarray}
\Psi(\rho,\phi,z,t) & = & e^{-k_{I 0} z}e^{ik_{R 0} z}e^{-i \om_0
t} \nonumber \\
 & \times & \left[
\frac{\exp\left(\dis{-q^2\frac{\rho^2}{\mu}}\right)}{\mu}
        \exp\left(\dis{- i k(\om_0) \frac{z}{\mu}}\right)\sum^{N}_{m=-N}A_{m}
        J_{\nu}\left(\eta_m\frac{\rho}{\mu}\right)e^{i\nu\phi}\,
        \exp\left(\dis{i \zeta_m \frac{z}{\mu}}\right) \right] \nonumber \\
 & \times & \left[ \exp\left[\frac{6a\left(2\varepsilon\tau-Z^{2}\right)+
iZ\left(-6a^{2}-6\varepsilon\tau+Z^{2}\right)}{12}\right]
\mathrm{Ai}\left(\varepsilon\tau-\frac{Z^{2}}{4}-iaZ\right) \right] \,\, ,
\label{psi2}
\end{eqnarray}

with $\mu$, $q$, $\eta_m$, $\zeta_m$ and $A_m$ given by
Eqs.(\ref{mu},\ref{q2},\ref{etan},\ref{zetam},\ref{An2}), being
that the parameter $a$ can be estimated from Eq.(\ref{a}) once we
have chosen the desired distance ($Z_{disp}$) of dispersion
resistance for the pulse.

\h The new pulse solutions represented by Eq.(\ref{psi2}),
propagating in dispersive, absorbing and unguided media and
carrying OAM, are resistant to the concomitant effects of
diffraction, dispersion and attenuation for long distances.
Actually, we can construct such pulses in such a way that we can
choose where and how intense their peaks will be within the
longitudinal spatial range $0 \leq z \leq L/2$.

\h A natural question that arises is about the generation process of these new
pulses. Interestingly, this can be done in a relatively simple way, through a
combination of the FW-type beams generation techniques with those for the generation
of Airy-type pulses. In this sense, through a Gaussian frequency spectrum with a
cubic phase \cite{nature}, an Airy-Gaussian pulse is created and, thereafter, it
is addressed onto a spatial light modulator, which encodes on it the FW's
hologram transmission function. The desired resulting pulse can then be obtained
after a 4f optical system and an iris.

\section{Examples}


\h In this section we will present two examples obtained from our theoretical method.
The material medium considered here is a SF10 glass and the central (free space)
wavelength used here is $\lambda_0 = 454.6$nm, which corresponds to an angular
frequency $\om_0 = 4.1464 \times 10^{15}$Hz.

\h At this frequency, the SF10 glass presents the following optical constants:
refractive index $n = n_R + in_I$, with $n_R=1.7554$ and $n_I= 1.5051 \times 10^{-7}$;
absorption coefficient $\alpha = 0.0416 {\rm cm}^{-1}$, which implies an
attenuation length $L_{att} = 1/\alpha = 24$cm; $\beta_1 = 6.3383 \times 10^{-9}$s/m,
which implies $ v_g=1.5777 \times 10^8 $m/s; dispersion coefficient
$\beta_2 = 459.73 {\rm fs}^2/$mm.

\h The following two examples deal with a vortex and a non-vortex pulse,
respectively, both with $\varepsilon=-1$ and a main lobe of duration $T_0 = 200$fs,
which implies a dispersion length $L_{disp}=T_0^2/|\beta_2| = 8.7$cm. In the first
case, the vortex pulse of topological charge $\nu=4$ has a donut shape of radius
approximately $\rho_4 =49\mu$m, which implies a Rayleigh distance (diffraction
length) $L_{diff} \approx 10$cm; in the second case, the non-vortex pulse has a
spot radius $r_0 = 22.1 \mu$m, which implies a Rayleigh distance
$L_{diff} = \sqrt{3}k_{R0}r_0^2 \approx 2$cm. The resulting pulses will be
designed to resist the effects of diffraction, dispersion and attenuation
till $z = 87$cm.


\subsection{First case}

\h Let us use our solution given by Eq.(\ref{psi2}), with $\varepsilon=-1$, to
construct a diffraction-dispersion-attenuation
resistant vortex pulse with topological charge $\nu = 4$, radius of the
donut-shaped-pulse $\rho_4 \approx 49 \mu$m, main lobe of duration $T_0 = 200$fs
and whose peak's intensity pattern along the propagation is given by a on-off-on
pattern, i.e, it is dictated by $|F(z)|^2$ with:

\bb F(z) \ug [H(z - \el_1)-H(z-\el_2)] + [H(z-\el_3)-H(z-\el_4)] \,\,\, , \label{F1} \ee
where $H(.)$ is the Heaviside function and $\el_1=0$, $\el_2=29$cm, $\el_3=58$cm
and $\el_4=87$cm. To this case, we can choose $L=1.74$m and, according to the
desired donut-shaped-pulse's radius, the value of the parameter $Q$ results to
be $Q = 0.99999 n_R\om/c$. The morphological function $F(z)$, besides pre-defining
the pulse's peak intensity behavior, also defines the diffraction-attenuation
resistance distance\footnote{According to our method, such characteristics are
acquired by the resulting pulse, $\Psi(\rho,\phi,z,t) = W(\rho,\phi,z)\,P(z,T)$,
through the function $W(\rho,\phi,z)$.}, which in this case is $87$cm. To get
dispersion resistance for such a distance, we have to set $a=0.02$. Now, the
resulting pulse (\ref{psi2}) can be completely characterized through
Eqs.(\ref{q2}-\ref{G}). In this case, we use $N=60$.

\h Figure 1 shows the comparison between the desired intensity pattern for the
pulse's peak along the propagation, given by
$|F(z)|^2$, and $|\Psi(\rho=\rho_4, \phi, z, t = t(z)) |^2$, which is the evolution
of the resulting pulse's peak intensity, which occurs, approximately,
on $ \rho = \rho_4 = 49 \mu$m and at times given by
$t = -(\beta_2^2/4T_0^3)z^2 + z/v_g + 1.019 T_0$ for the a given position
$z$ of the pulse's peak. We can see a good agreement between them.

\begin{figure}[!h]
\begin{center}
 \scalebox{.35}{\includegraphics{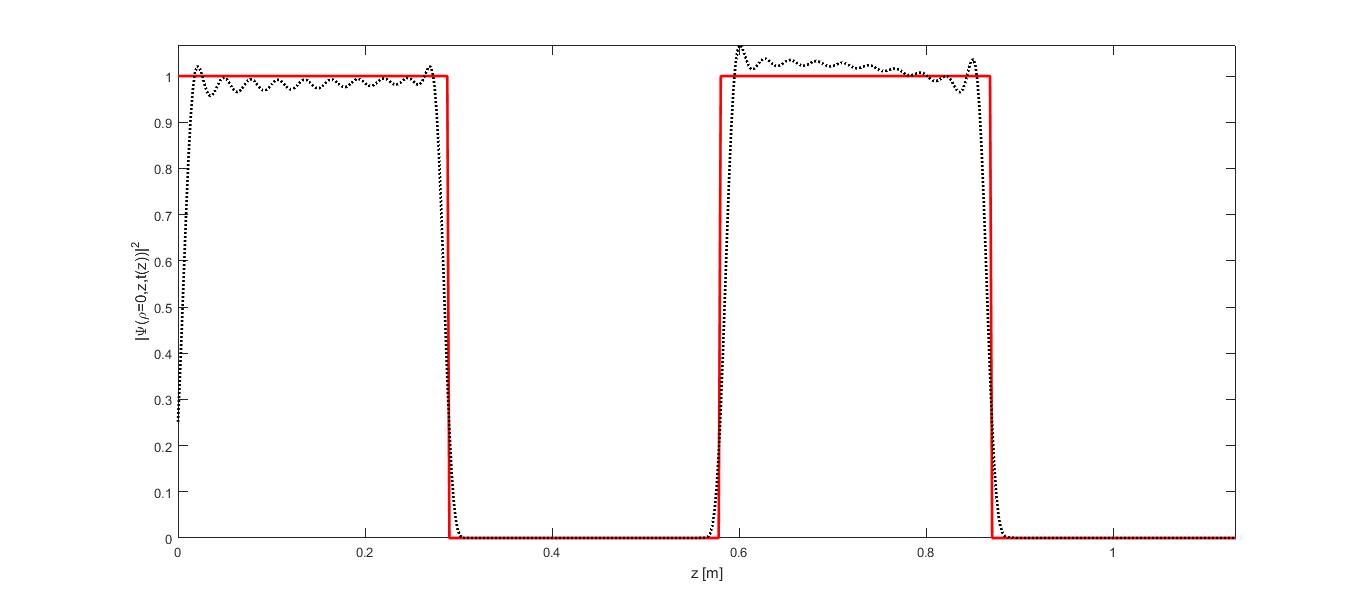}}
\end{center}
\caption{Comparison between the desired intensity pattern for the pulse's peak
along the propagation (continuous red line), given by $|F(z)|^2$,
and $|\Psi(\rho=\rho_4, \phi, z, t = t(z)) |^2$, which is the evolution
of the resulting pulse's peak intensity (doted line), which occurs,
approximately, on $ \rho = \rho_4 = 49 \mu$m and at
times $t = -(\beta_2^2/4T_0^3)z^2 + z/v_g + 1.019 T_0$ for the a given position
 $z$ of the pulse's peak. We can see a good agreement between them.
} \label{fig1}
\end{figure}

\h Figure 2 shows the temporal evolution of the resulting pulse at different
distances (i.e., values of $z$) and considering $\rho = \rho_4 = 49 \mu$m.
It is evident that the pulse intensity obeys the required on-off-on behavior
and also that it preserves the temporal width of its main lobe till the distance
pre-defined by the morphological function $F(z)$.

\begin{figure}[!h]
\begin{center}
 \scalebox{.35}{\includegraphics{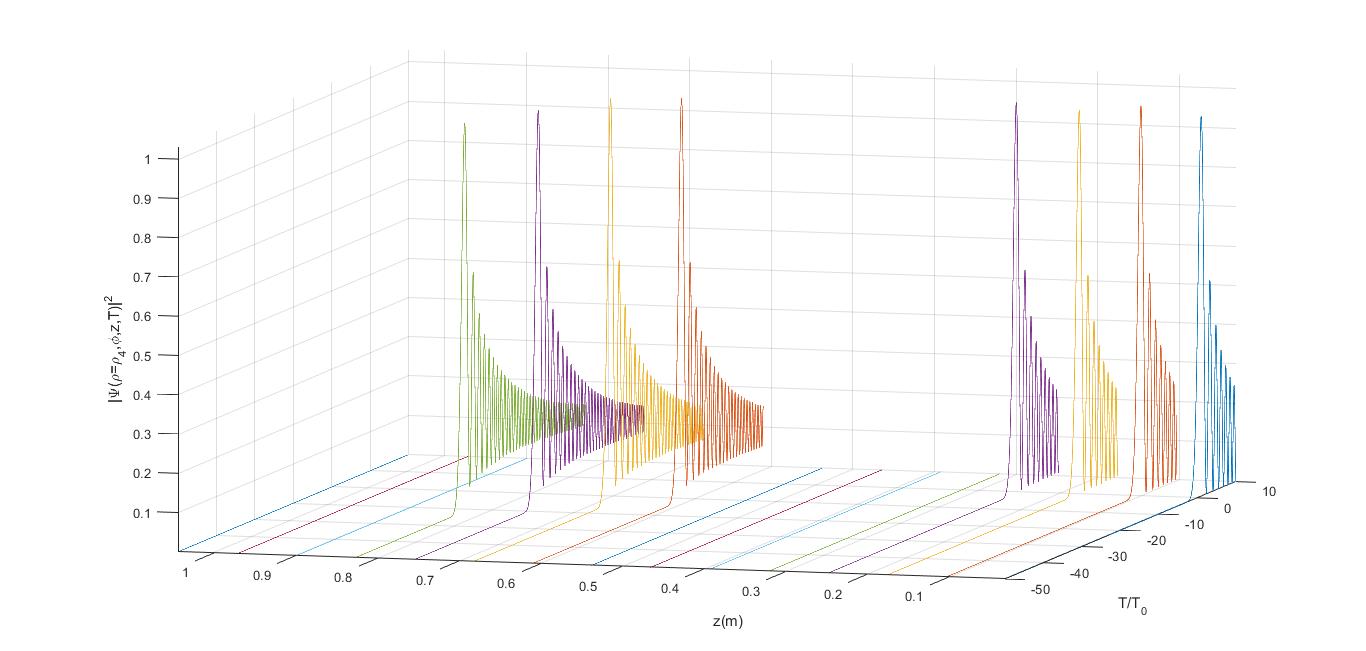}}
\end{center}
\caption{Temporal evolution of the resulting pulse at different distances and
considering $\rho = \rho_4 = 49 \mu$m. One can see that the pulse intensity
obeys the required on-off-on behavior and also that it preserves the temporal
width of its main lobe till a distance 10 times greater than the dispersion
length of an ordinary pulse with the same temporal width.} \label{fig2}
\end{figure}

\h Finally, Fig.3 shows the $3D$ pulse intensity, $|\Psi(\rho,\phi,z,t)|^2$,
at nine
different instants of time. The first, second and third lines of the subfigures
show the pulse evolution within
the ranges $\el_1<z<\el_2$, $\el_2<z<\el_3$ and $\el_3<z<\el_4$, respectively.
We can see that, in addition to the vortex pulse having the desired space-time
evolution, it is resistant to the effects of the diffraction, dispersion and absorption.

\begin{figure}[!h]
\begin{center}
 \scalebox{.37}{\includegraphics{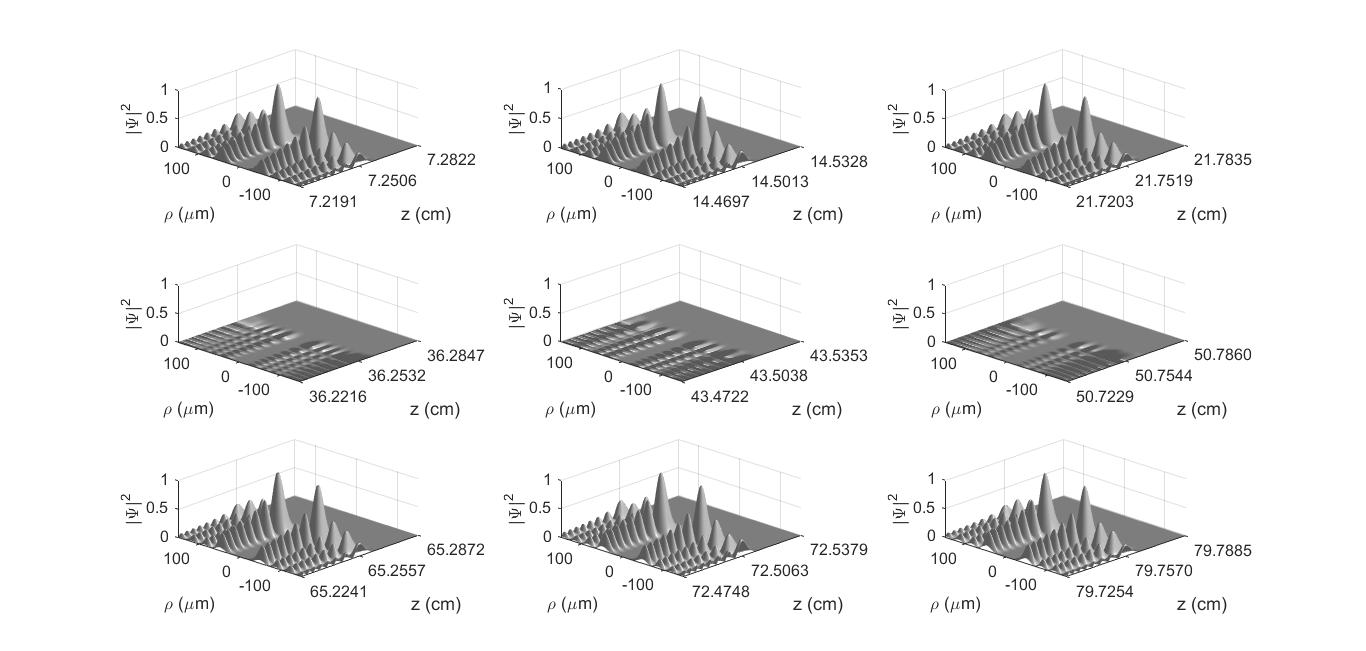}}
\end{center}
\caption{The $3D$ pulse intensity, $|\Psi(\rho,\phi,z,t)|^2$, of the first example
at nine
different instants of time. The first, second and third lines of the subfigures
show the vortex pulse evolution within
the ranges $\el_1<z<\el_2$, $\el_2<z<\el_3$ and $\el_3<z<\el_4$, respectively.
We can see that, in addition to the vortex pulse having the desired space-time
evolution, it is resistant to the effects of the diffraction, dispersion and
absorption.} \label{fig3}
\end{figure}

\subsection{Second case}

\h Here, we use our solution, Eq.(\ref{psi2}), to construct a
diffraction-dispersion-attenuation
resistant pulse with null topological charge, i.e. $\nu = 0$, spot radius
$r_0 \approx 22 \mu$m, main lobe of duration $\tau_0 = 200$fs and whose
peak's intensity pattern along the propagation is given by a ladder
pattern, i.e, it is dictated by
$|F(z)|^2$ with:

\bb \begin{array}{clr} F(z) &=
\sqrt{1}\,[H(z-\el1)-H(z-\el2)]\\
\\
&+ \sqrt{2}\,[H(z-\el2)-H(z-\el3)] \\
\\
&+ \sqrt{3}\,[H(z-\el3)-H(z-\el4)]
\,\, ,
\end{array} \label{F2} \ee

where $H(.)$ is the Heaviside function and, as before, $\el_1=0$, $\el_2=29$cm,
$\el_3=58$cm and $\el_4=87$cm. To this case, we can again choose $L=1.74$m and,
according to the desired pulse spot radius, the value of the parameter $Q$
results to be $Q = 0.99999 n_R \om/c$. The morphological function $F(z)$,
besides pre-defining the pulse's peak intensity behavior, also defines the
diffraction-attenuation resistance distance, which in this case is $87$cm.
For obtaining dispersion resistance for such a distance, we have to set
$a=0.02$ and the resulting pulse (\ref{psi2}) can be completely characterized
through Eqs.(\ref{q2}-\ref{G}).

\h Figure 4 compares the desired intensity pattern for the pulse's peak along
the propagation, given by $|F(z)|^2$, with
$|\Psi(\rho=0,z, t = t(z))|^2$, the evolution of the resulting pulse's peak
intensity, which occurs on $\rho = 0$ and at
times $t = -(\beta_2^2/4T_0^3)z^2 + z/v_g + 1.019 T_0$ for the a given
position $z$ of the pulse's peak. We can see a good agreement between them.

\begin{figure}[!h]
\begin{center}
 \scalebox{.55}{\includegraphics{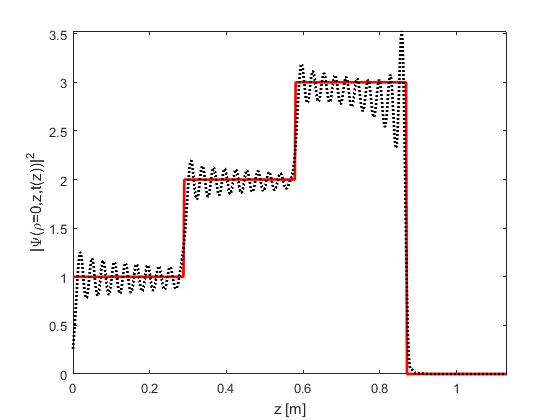}}
\end{center}
\caption{Comparison between the desired intensity pattern for the pulse's peak
along the propagation (continuous red line), given by $|F(z)|^2$,
with $|\Psi(\rho=0,z, t = t(z))|^2$, the evolution of the resulting pulse's
peak intensity (doted line), which occurs on $\rho = 0$ and at times
$t = -(\beta_2^2/4T_0^3)z^2 + z/v_g + 1.019 T_0$ for the a given position
$z$ of the pulse's peak. We can see a good agreement between them} \label{fig4}
\end{figure}

\h The on-axis ($\rho=0$) temporal evolution of the resulting pulse at
different distances is shown in Fig.5. It is very clear that the pulse
intensity obeys the required ladder behavior, preserving the temporal
width of its main lobe till the distance pre-defined by the morphological
function $F(z)$.

\begin{figure}[!h]
\begin{center}
 \scalebox{.35}{\includegraphics{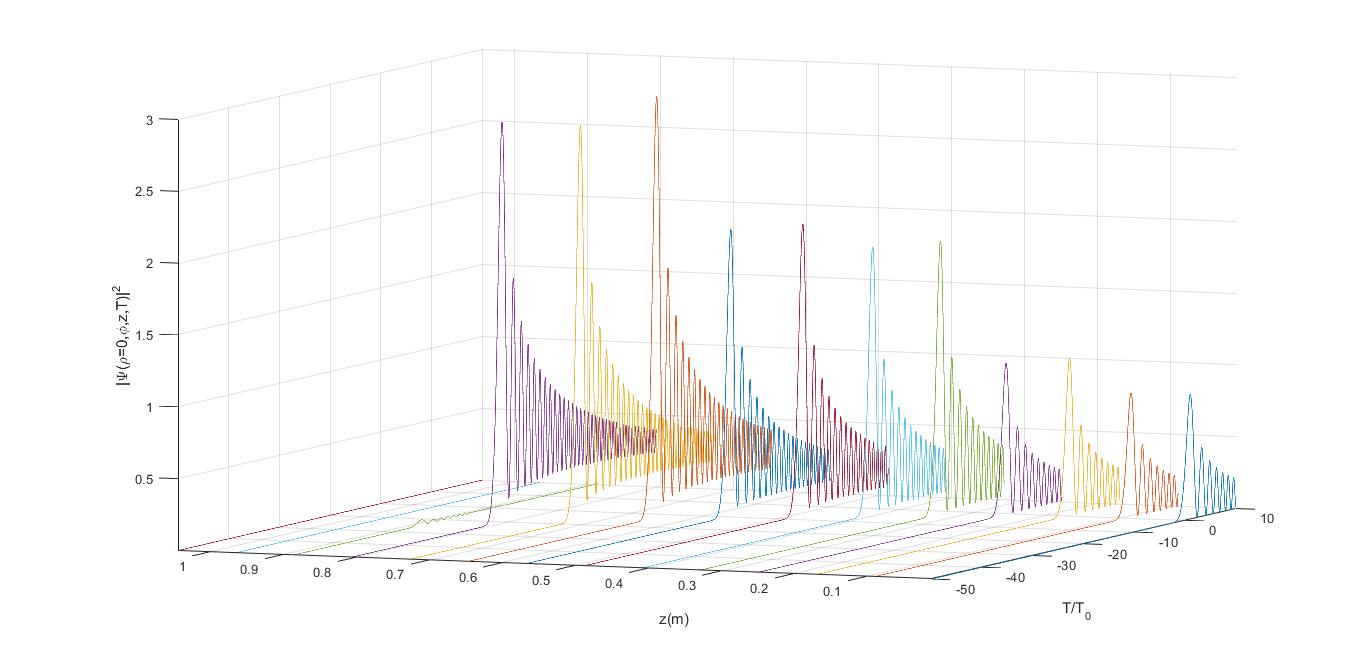}}
\end{center}
\caption{On-axis ($\rho=0$) temporal evolution of the resulting pulse at
different distances. One can see that the pulse intensity obeys the required
ladder behavior, preserving its temporal width till a distance 10 times
greater than the dispersion length of an ordinary pulse with the same temporal
width.} \label{fig5}
\end{figure}

\begin{figure}[!h]
\begin{center}
 \scalebox{.37}{\includegraphics{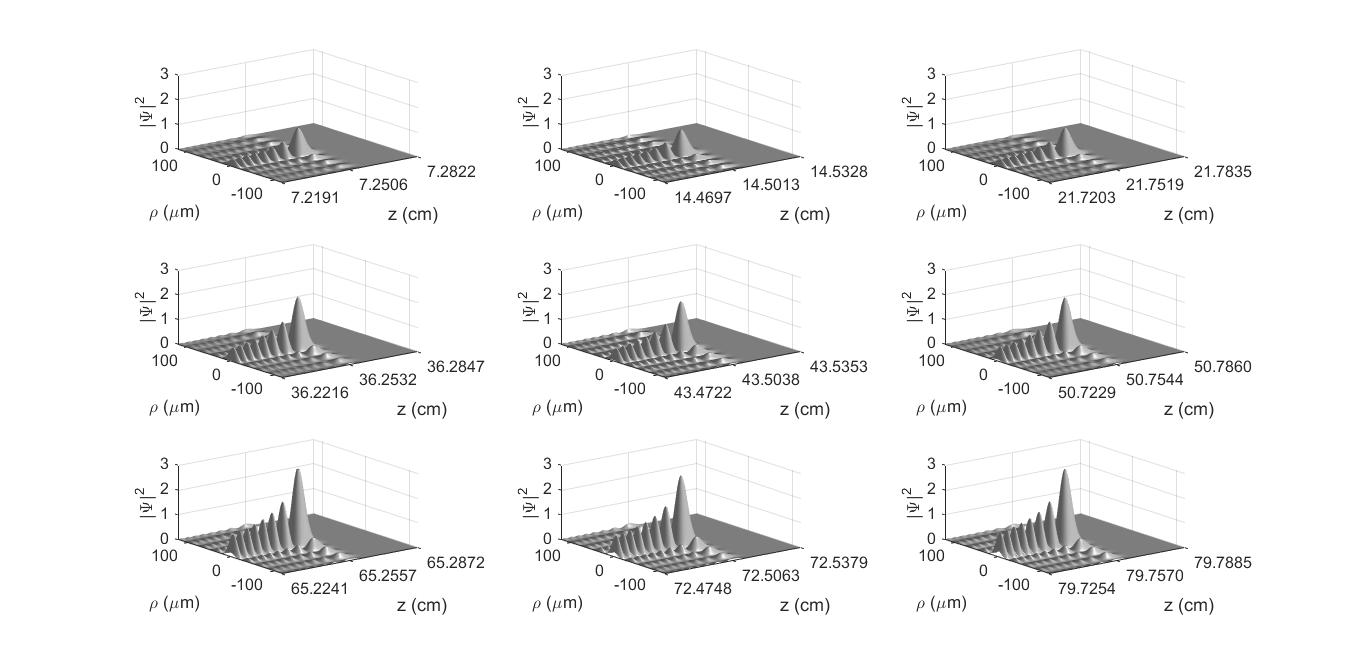}}
\end{center}
\caption{The $3D$ pulse intensity, $|\Psi(\rho,z,t)|^2$, at nine
different instants of time. The first, second and third lines of the subfigures
show the pulse evolution within
the ranges $\el_1<z<\el_2$, $\el_2<z<\el_3$ and $\el_3<z<\el_4$, respectively.
In addition to the desired pulse's space-time evolution, one can see the pulse
is resistant to the concomitant effects of diffraction, dispersion and
absorption.} \label{fig6}
\end{figure}

\h Finally, Fig.6 depicts the $3D$ pulse intensity, $|\Psi(\rho,z,t)|^2$, at nine
different instants of time. The first, second and third lines of the subfigures
show the pulse evolution within
the ranges $\el_1<z<\el_2$, $\el_2<z<\el_3$ and $\el_3<z<\el_4$, respectively.
We can see, in addition to the desired pulse's space-time evolution, it is
resistant to the concomitant effects of diffraction, dispersion and absorption.

\section{Conclusions}

In this work we have developed a method capable of providing, in unguided dispersive and absorbing media, vortex pulses resistant to the three concomitant effects of diffraction, dispersion and attenuation.  As a matter of fact, with our approach it is possible to perform a space-time modelling on such new pulses, i.e., it allows the choice of multiple spatial ranges where the pulse intensities can be chosen a priori.

Such approach is a result of a fusion between two important theoretical methodologies, one related to the so called Frozen-Wave-beams, which are non-diffracting beams whose spatial intensity pattern can be chosen \emph{a priori} in
absorbing media, the other related to the Airy-Type pulses, which
are pulses resistant to the dispersion effects in material dispersive media.

The new kind of pulses can have potential applications in
different fields as photonics, nonlinear optics, optical
communications, optical tweezers, optical atom guiding, medicine,
etc..

\section*{Acknowledgements}
Thanks are due to partial support from FAPESP (under grant 2015/26444-8) and from CNPq
(under grant 304718/2016-5).

\end{document}